\documentclass[aps,preprint,showpacs,groupedaddress]{revtex4}

\usepackage{graphicx}

\newcommand{\Tr}{\text{Tr}}
\newcommand{\ket}[1]{|#1\rangle}
\newcommand{\bra}[1]{\langle#1|}

\newcommand{\sz}{\sigma_z}

\newcommand{\sigmam}{\sigma^-}
\newcommand{\sigmap}{\sigma^+}
\newcommand{\ext}[2]{|#1\rangle\langle#2|}   %外积

\newcommand{\dt}[1]{\frac{d}{d t}#1} %对t的偏导数

\newcommand{\PRA}[3] {Phys. Rev. A {\bf #1}, #2
(#3)}
\newcommand{\PRB}[3] {Phys. Rev. B {\bf #1}, #2
(#3)}
\newcommand{\PRL}[3] {Phys. Rev. Lett. {\bf #1}, #2
(#3)}

\newcommand{\EPL}[3] {Europhysics Letters {\bf #1}, #2 (#3)}
\newcommand{\PS}[3] {Phys. Scr. {\bf #1}, #2 (#3)}

\begin{document}

%\draft
\title{Non-Markovian Quantum Jump with Generalized Lindblad Master
Equation }
\author{X. L. Huang\footnote{ghost820521@163.com}}
\author{H. Y. Sun}
\author{X. X. Yi\footnote{yixx@dlut.edu.cn}}
\affiliation{School of physics and optoelectronic technology,\\
Dalian University of Technology, Dalian 116024 China }

\date{\today}

\begin{abstract}
The Monte Carlo wave function method or the quantum trajectory/jump
approach is a powerful tool to study dissipative dynamics governed
by the Markovian master equation, in particular for high-dimensional
systems and when it is  difficult  to simulate directly. In this
paper, we extend this method to the non-Markovian case described by
the generalized Lindblad master equation. Two examples to illustrate
the method are presented and discussed. The results show that the
method can correctly reproduce the dissipative dynamics for the
system. The difference between this method and  the traditional
Markovian jump approach  and the computational efficiency of this
method are also discussed.

\end{abstract}

 \maketitle

\section{Introduction}
Since the  pioneering work of Albert Einstein, who explained the
phenomenon of dissipation and Brownian motion in his annus mirabilis
of 1905 by use of statistical methods, a rich variety of methods to
tackle quantum fluctuations and quantum dissipation in open systems
has been proposed\cite{quantumnoisy,dissipation}. Among them  the
 quantum master equation (QME) approach and  the quantum
Langevin description (QLE)\cite{quantumoptics} are two of  powerful
functional integral techniques for the study of  time evolution of
open quantum systems. The quantum master equation can be divided
into two categories: Markovian and non-Markovian.  The Markovian
master equation\cite{Lindblad} (especially in the Lindbald form) can
be derived with the weak coupling limit(or the Born approximation)
and the Markovian approximation. It can be solved
analytically\cite{solvemasterequ} for some special cases, but for
most cases we have to  solve and simulate it  numerically by the
Monte Carlo wave function method or quantum trajectory/jump
approach\cite{wavefunction,geometric,IJQI,Berryphaseyi,
superdensecoding,Teleportation}. This method is very effective for
 qubit systems even with large number of qubits, say
$n=24$\cite{Teleportation}.

However, the dynamics of an open system is not always Markovian.
Strong system-environment couplings, correlation and entanglement in
the initial state and structured reservoirs may lead the dynamics
far from Markovian. Many methods have been proposed to describe the
non-Markovian process, including  the Lindblad equation with time
dependent decay rates\cite{BPbookprojection}, generalized Lindblad
equation\cite{projection} obtained from the correlated projection
superoperator techniques\cite{projection3,projection2008},
phenomenological memory kernel master
equation\cite{expmemoryhazard,expmemorypositivity} and the
post-Markovian master
equation\cite{postMarkovian,postMarkoviansolutionPRA,postMarkoviansolutionHuang}.
The first two methods are local in time  while the last two involve
an integral of time. For the first method, the only  difference from
the Markovian master equation is that the decay rates in the
equation are time-dependent. These decay rates  may take not only
positive values but also negative ones. When decay rates are
positive, the Markovian Monte Carlo wave function method can
directly be used. However, the mehtod is not available when the
decay rates are negative. This problem was solved in
Ref.\cite{nonMQJ} by introducing reversed jumps.

The generalized Lindblad master equation can well describe the
dynamics of an open system beyond the Markovian limit, especially it
is very effective for an environment composed of
spins\cite{spinbath1,spinbath2,spinbath3} and  structured
reservoirs\cite{projectionmodel}. However the extension of the Monte
Carlo simulation to this equation remains untouched. In this paper,
we will explore the unraveling and quantum trajectory approach for
the generalized Lindblad equation. The structure of this paper is
organized as follows. In Sec.\ref{sectionGeneralizedLME} we briefly
review the generalized Lindblad equation. In
Sec.\ref{sectionQuantumJump} we give the unraveling of this equation
and generalize the  Monte Carlo method to this equation.   Two
examples are presented  in Sec.\ref{sectionApplication}. Finally, we
conclude our results in Sec.\ref{sectionConclusion}.

\section{Generalized Lindblad Master
Equation\label{sectionGeneralizedLME}} The equation that governs the
dynamics of an open quantum system can be derived  by means of the
projection superoperator technique\cite{BPbookprojection,
projection3}. The form (Markovian or non-Markovian) of the master
equation crucially depends on the approximation used in the
derivation, reflecting in the projection superoperator chosen. When
we project the total system state into a tensor product, we can
obtain the Markovian master equation, whereas  a non-Markovian
master equation can be obtained when we use a correlated projection.
The following is the master equation derived by this method and it
is called  the generalized Lindblad master
equation\cite{projection},
\begin{eqnarray}
\dt
\rho_m=-i[H_m,\rho_m]+\sum_{n\lambda}\left(R_{mn}^{\lambda}\rho_nR_{mn}^{\lambda\dag}-
\frac12\{R_{nm}^{\lambda\dag}R_{nm}^{\lambda}, \rho_m\} \right),
\label{Gmasterequation}
\end{eqnarray}
where $H_m$ are Hermitian operators and $R_{mn}^{\lambda}$ are
arbitrary system operators depending on the form of
system-environment interactions. If we  have only a single component
$\rho_S=\rho_1$, this equation obviously reduces to the ordinary
Markovian master equation. In this paper we will focus on the case
where we have at least two components. The state of the reduced
system in this case is $\rho_S=\sum_m\rho_m$, we remind  that
$\Tr\rho_m<1$.

\section{Quantum Jump\label{sectionQuantumJump}}
For  clarity, we define the jump operators
$W_{mn}^{\lambda}=R_{mn}^{\lambda}$ and non-jump operators
$W_{mm}^0=I-i\mathcal{H}_mdt$, where the non-Hermitian effective
Hamiltonian is given by
$\mathcal{H}_m=H_m-\frac12i\sum_{n\lambda}R_{nm}^{\lambda\dag}R_{nm}^{\lambda}$.
There are two subscripts and one superscript for the operator
$W_{mn}^{\lambda}$. The first subscript $m$ denotes the index of
component where the system in,  while the second subscript $n$
denotes the index of component for the operation acting on; the
superscript $\lambda$ represents the jump mode. Initially we assume
that each operator $\rho_m(t_0)$ can be written as
$\rho_m(t_0)=\ext{\psi_m(t_0)}{\psi_m(t_0)}$, where
$\ket{\psi_m(t_0)}$ is a non-normalized wave function. After an
infinitesimal time $dt$, it evolves into the following state
\begin{eqnarray}
\rho_m(t_0+dt)=\sum_{n\lambda}\ext{\psi_{mn}^{\lambda}}{\psi_{mn}^{\lambda}}
dp_{mn}^{\lambda}+\ext{\psi_{mm}^0}{\psi_{mm}^0}dp_{mm}^0,\label{unraveling}
\end{eqnarray}
where the new states are defined by
\begin{eqnarray}
\ket{\psi_{mn}^{\lambda}}=\frac{\sqrt{p_m}W_{mn}^{\lambda}\ket{\psi_n(t)}}
{\left\|W_{mn}^{\lambda}\ket{\psi_n(t)}\right\|}, \label{statejump}
\end{eqnarray}
and
\begin{eqnarray}
\ket{\psi_{mm}^0}=\frac{\sqrt{p_m}W_{mm}^0\ket{\psi_m(t)}}{\left\|W_{mm}^0\ket{\psi_m(t)}\right\|},
\label{statenonjump}
\end{eqnarray}
with probabilities
\begin{eqnarray}
&&dp_{mn}^{\lambda}=\frac1{p_m}\bra{\psi_n(t_0)}W_{mn}^{\lambda\dag}W_{mn}^{\lambda}\ket{\psi_n(t_0)}dt,\nonumber\\
&&dp_{mm}^0=\frac1{p_m}\bra{\psi_m(t_0)}W_{mm}^{0\dag}W_{mm}^0\ket{\psi_m(t_0)},
\label{probability}
\end{eqnarray}
respectively.  In Eqs.(\ref{statejump}) and (\ref{statenonjump}),
\begin{eqnarray}
p_m=\sum_{n\lambda}\bra{\psi_n(t_0)}W_{mn}^{\lambda\dag}W_{mn}^{\lambda}\ket{\psi_n(t_0)}dt+
\bra{\psi_m(t_0)}W_{mm}^{0\dag}W_{mm}^0\ket{\psi_m(t_0)},\label{probabilitytotal}
\end{eqnarray}
is the weight for the component $\rho_m$ that  satisfies
\begin{eqnarray}
p_m=\Tr\rho_m(t+dt).
\end{eqnarray}
Note that the jumps for $\rho_m$ depend on the other components
$\rho_n, (n\neq m)$ of the reduced density  matrix $\rho$. This
makes our method different from the traditional quantum jump method.

We can prove this unraveling by taking the jump and non-jump states
(\ref{statejump}), (\ref{statenonjump}) and the probabilities
(\ref{probability}), (\ref{probabilitytotal}) into
Eq.(\ref{unraveling}),
\begin{eqnarray}
&&\rho_m(t+dt)\nonumber\\
&&=\sum_{n\lambda}W_{mn}^{\lambda}\ext{\psi_n(t)}{\psi_n(t)}W_{mn}^{\lambda\dag}dt+
W_{mm}^0\ext{\psi_m(t)}{\psi_m(t)}W_{mm}^{0\dag}.\label{unr}
\end{eqnarray}
Simple algebra shows that in  the limit $dt\rightarrow0$,
Eq(\ref{unr}) reveals Eq.(\ref{Gmasterequation}). The evolution
governed by Eq.(\ref{Gmasterequation}) can be simulated numerically
by the so-called Monte Carlo wave function approach according to the
unraveling given above. We start the time evolution from the state
$\rho(t_0)=\sum_m\rho_m(t_0)=\sum_m\ext{\psi_m(t_0)}{\psi_m(t_0)}$,
where $\rho_m (m=1,2,3,...)$ are the components for $\rho$. At time
$t_0+dt$, where $dt$ is much smaller than the timescale relevant for
the evolution of the density matrix, a random number $\epsilon$
which is  randomly distributed  in the unit interval $[0,1]$ is used
to determine the jump. Note that all the components are controlled
by this random number. For each component $\ket{\psi_m}$, if
$0\leq\epsilon\leq dp_{m1}^1$, it jumps to $\ket{\psi_{m1}^1}$, if
$dp_{m1}^1<\epsilon\leq dp_{m1}^1+dp_{m1}^2$, it jumps to
$\ket{\psi_{m1}^2}$, and so on. These jumps are all operated on the
component $\rho_1$; if
$\sum_{\lambda}dp_{m1}^{\lambda}<\epsilon\leq\sum_{\lambda}dp_{m1}^{\lambda}+dp_{m2}^1$,
it jumps to the  component $2$, namely $\ket{\psi_{m2}^1}$. Jumps to
the other components can be established in a similar way. If
$\epsilon>\sum_{n\lambda}dp_{mn}^{\lambda}$, a non-jump takes place
and the state ends up in $\ket{\psi_{mm}^0}$. This operation is
acted on the component $\rho_m$ itself. We define a generalized jump
superoperator $\mathcal {W}_i$, which denotes all  jumps for all the
components controlled by this random number. We repeat this process
as many times as $n=\Delta t/dt$ for all the components, where
$\Delta t$ is the total evolution time. We call this single
evolution a generalized quantum trajectory. This trajectory contains
all the components of the density matrix. Given an operator $A$, we
can write its mean value $\langle A\rangle(t)=\Tr(A\rho(t))$ as an
average over $\mathcal {N}$ trajectories as
\begin{eqnarray}
\langle A\rangle(t)=\lim_{\mathcal
{N}\rightarrow\infty}\sum_{j=1}^{\mathcal {N}} \sum_m
\bra{\psi_{m,j}(t)}A\ket{\psi_{m,j}(t)}.
\end{eqnarray}

\section{Application\label{sectionApplication}}

\begin{figure}
\includegraphics*[bb=1 1 230 195, width=6cm]{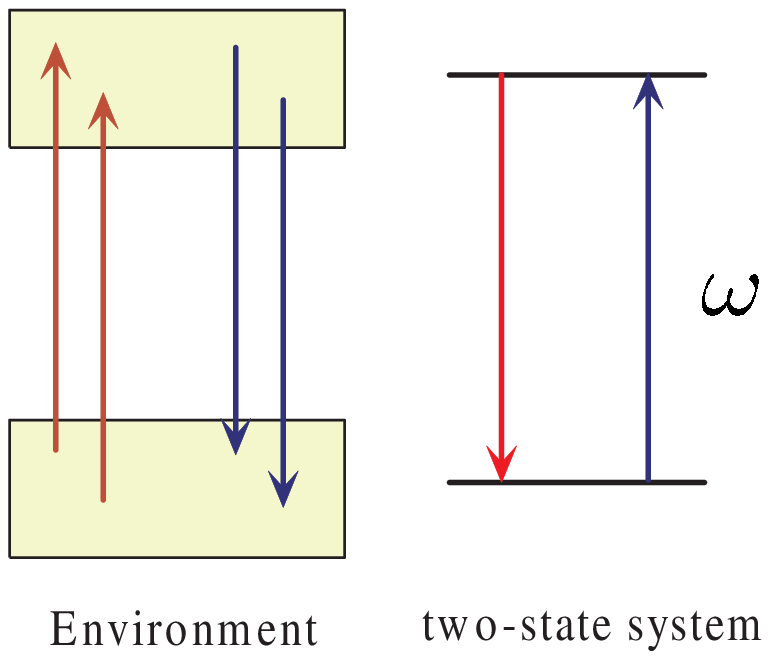}\caption{A two-state system coupled to an
environment consisting of two energy bands with a finite number of
levels.} \label{setup}
\end{figure}

\begin{figure}
\includegraphics*[width=0.8\columnwidth,
height=0.5\columnwidth]{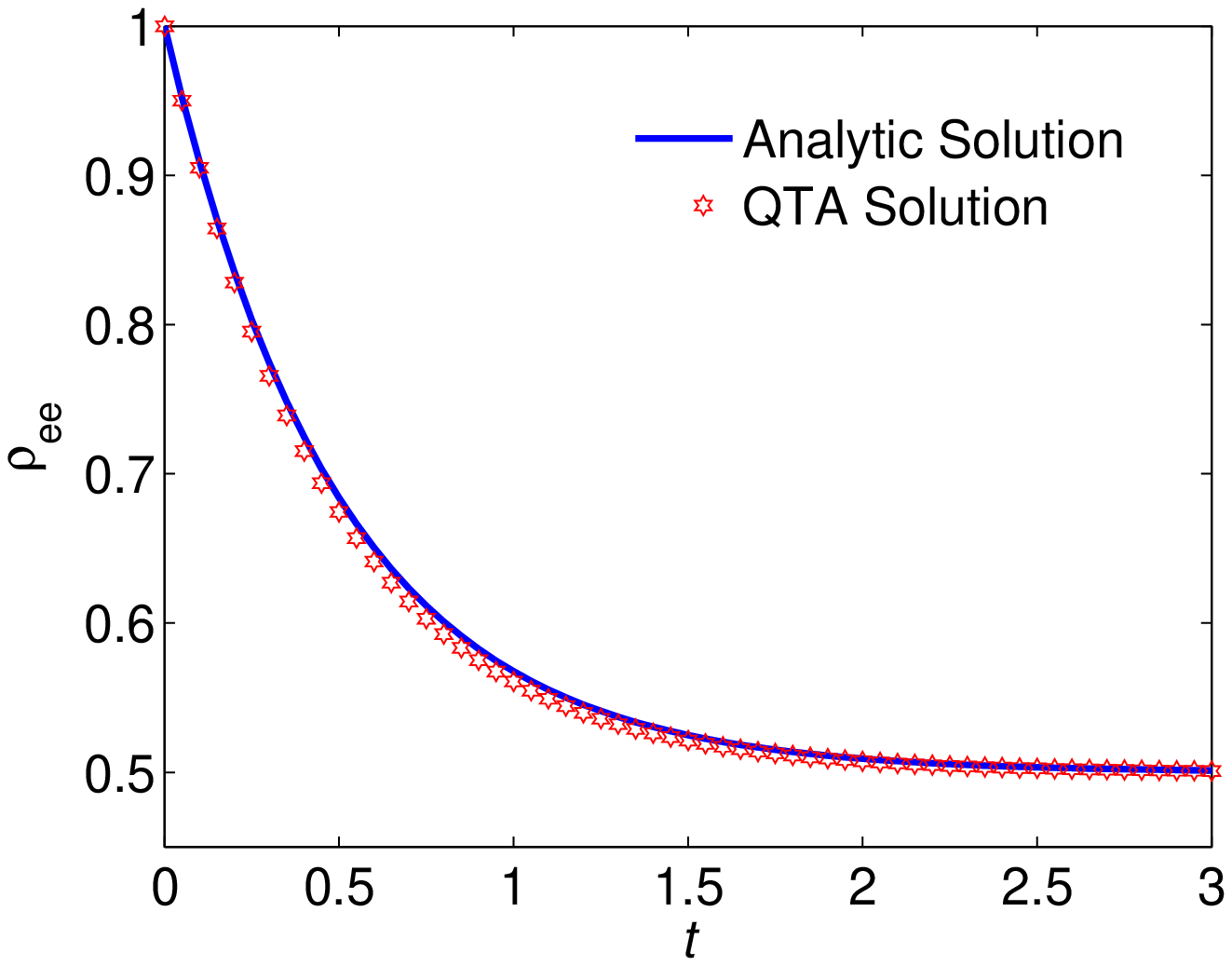}
\includegraphics*[width=0.8\columnwidth,
height=0.5\columnwidth]{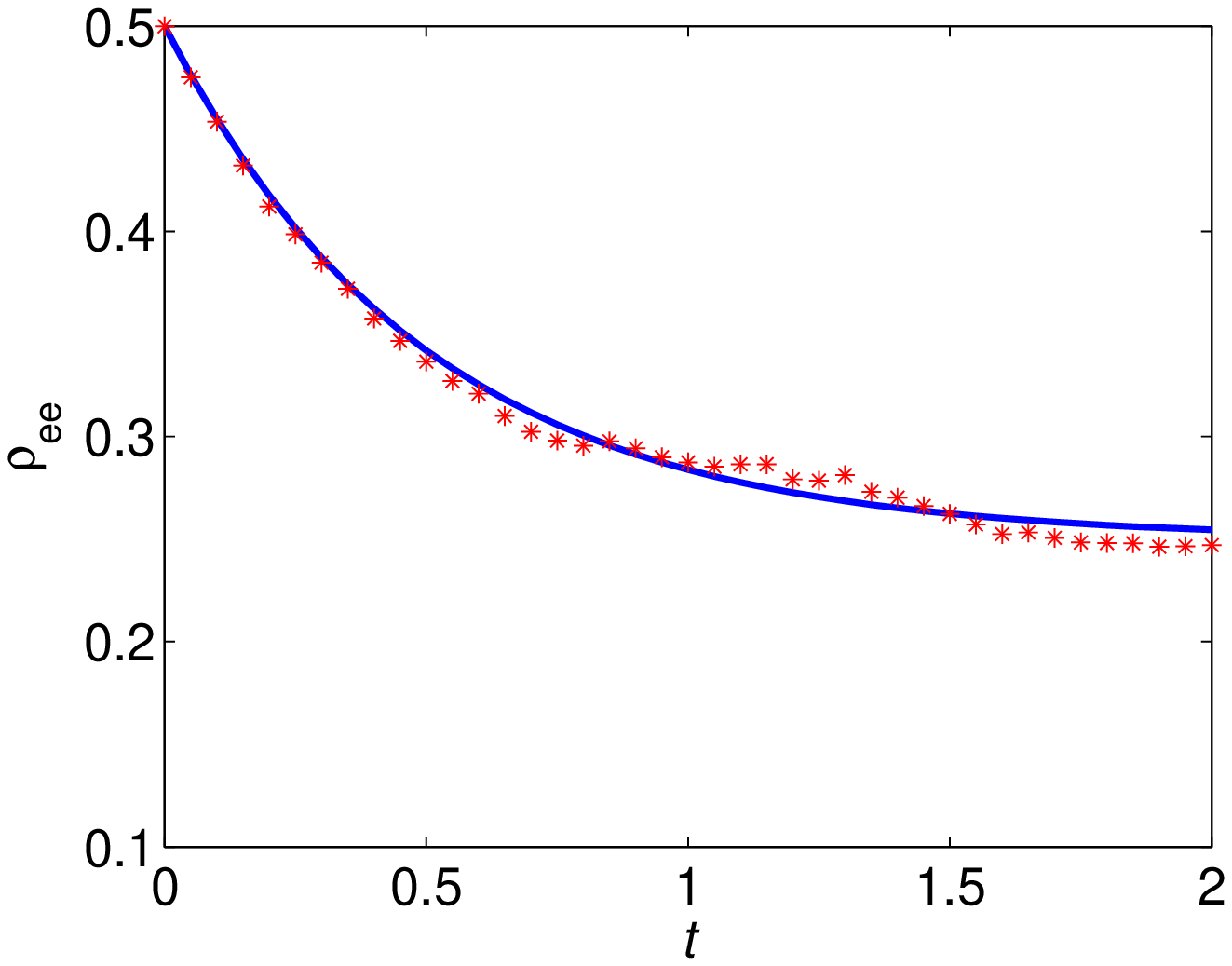} \caption{(Color online)
Comparisons of the analytic solution for Eq.(\ref{exaplemaster}) to
the results given by quantum trajectory approach. The initial state
of the system is chosen $\ket{\phi(0)}=\ket{e}$ in the top figure
while $\ket{\phi(0)}=\frac1{\sqrt2}(\ket{e}+\ket{g})$ in the bottom
one. Initially only the lower band of the environment is populated.
The other parameters chosen are $\gamma_1=\gamma_2=1$. The time $t$
is plotted in units of $1/\hbar$.} \label{c1}
\end{figure}

\begin{figure}
\includegraphics*[width=0.8\columnwidth,
height=0.5\columnwidth]{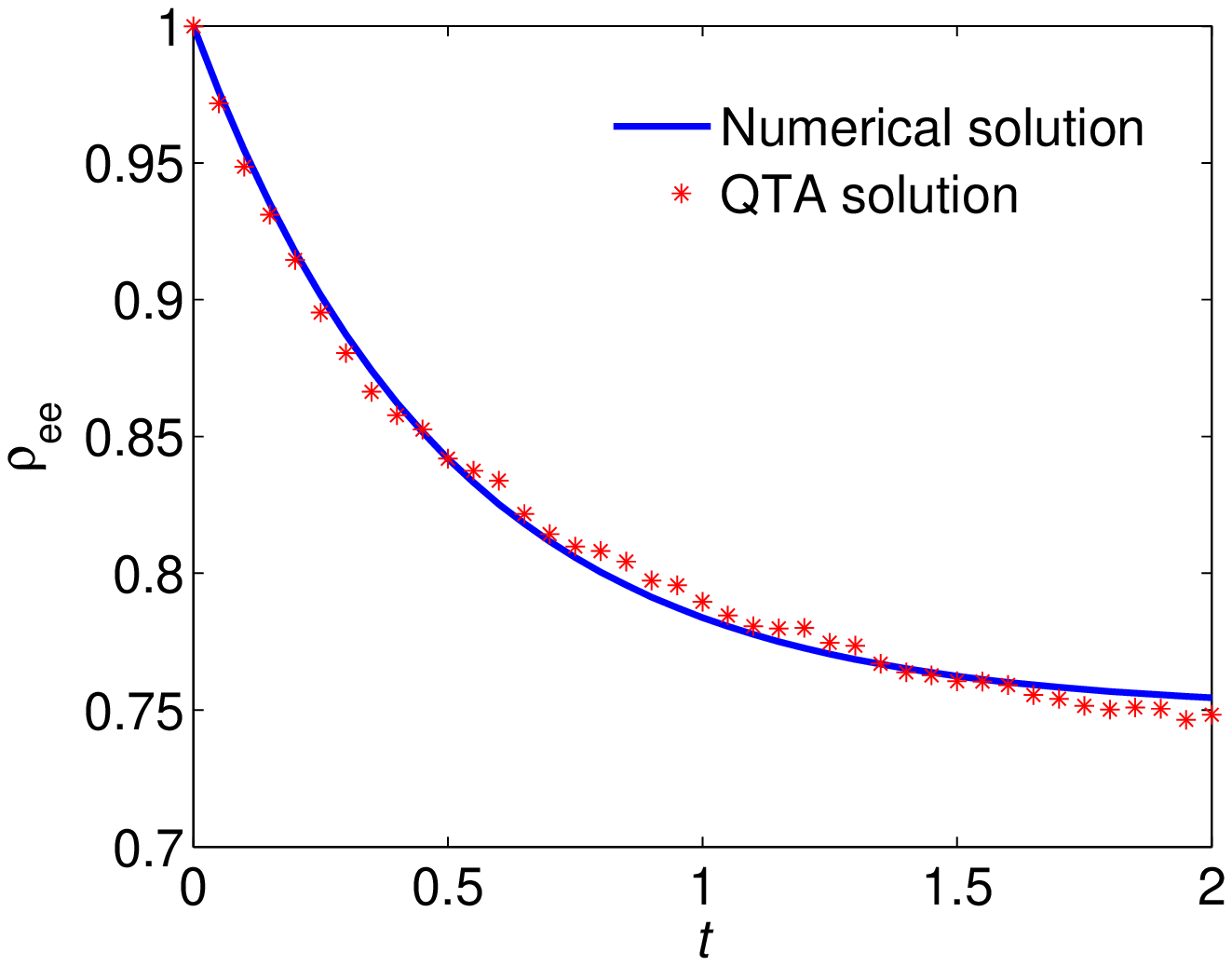}
\includegraphics*[width=0.8\columnwidth,
height=0.5\columnwidth]{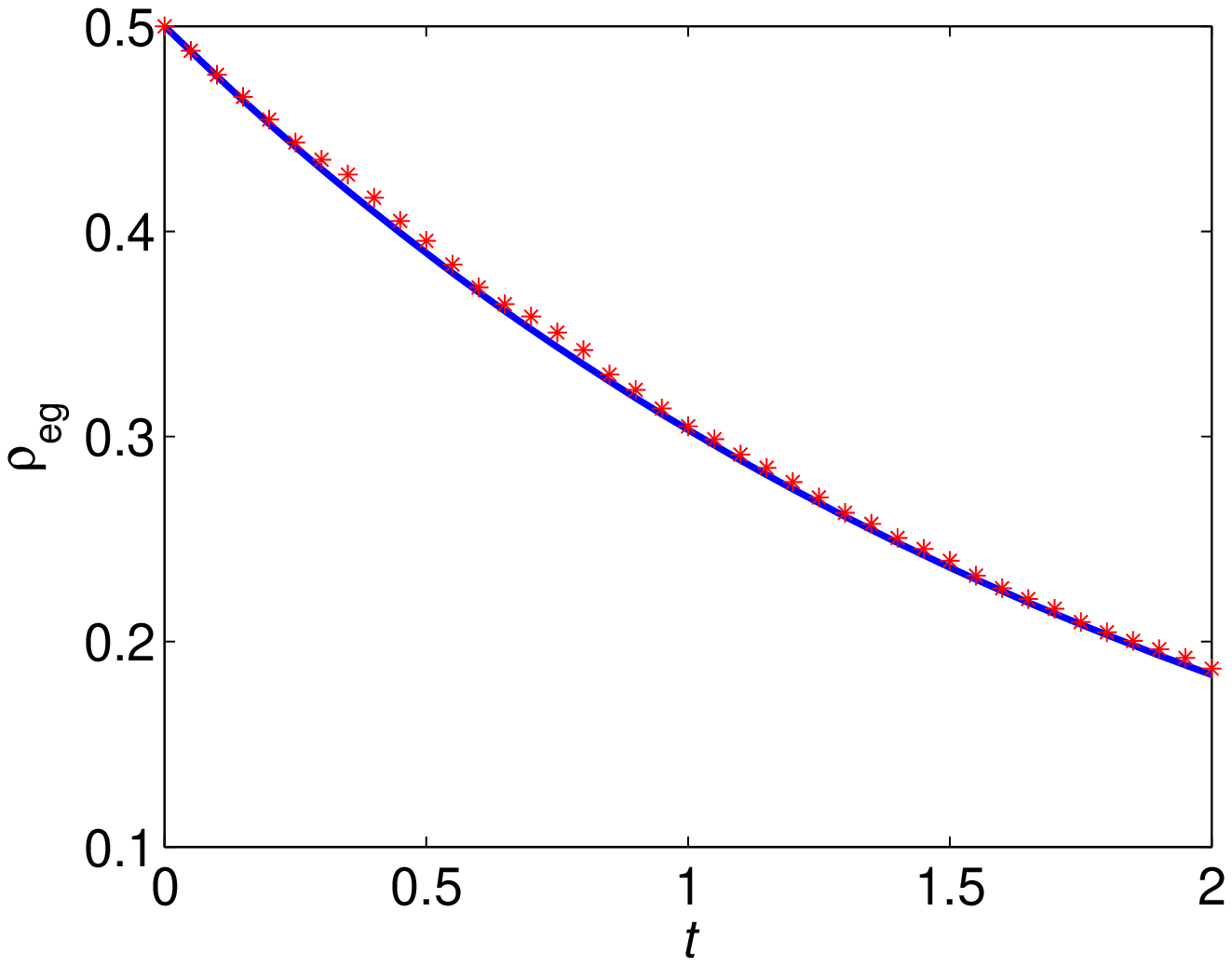} \caption{(Color online)
Comparisons of the numerical solution for Eq.(\ref{exaplemaster}) to
the results given by quantum trajectory approach. The initial state
of the system is chosen $\ket{\phi(0)}=\ket{e}$ in the top figure
while $\ket{\phi(0)}=\frac1{\sqrt2}(\ket{e}+\ket{g})$ in the bottom
one. Initially the two bands of the environment are populated. Two
parameters are $\gamma_1=\gamma_2=1$. The time $t$ is plotted in
units of $1/\hbar$. Note that in the bottom figure we plot the
off-diagonal element $\rho_{eg}$ of the reduced system.} \label{c2}
\end{figure}

In this section, we use the model and the generalized master
equation given in Refs.\cite{projectionmodel} and \cite{spinbath2}
as two examples to illustrate our method. First consider a two-state
system coupled to an environment. The environment consists of a
large number of energy levels which are arranged into two energy
bands with the same energy spacing(see Fig.\ref{setup}).  The lower
energy band contains $N_1$ levels while the upper one  $N_2$ levels.
This model can be understood  as a "many level" environment or
"container", of which only the relevant parts of the spectrum enter
the model. For details of this model, we refer the reader to
\cite{model2,model3}. The total Hamiltonian for a qubit coupled to
such an  environment in Schr\"{o}dinger picture is
 $H=H_0+V$\cite{projectionmodel} with(we set $\hbar=1$)
\begin{eqnarray}
H_0=\frac12\omega\sigma_z+\sum_{n_1}\frac{\delta\epsilon}{N_1}n_1|n_1\rangle\langle
n_1|+\sum_{n_2}(\omega+\frac{\delta\epsilon}{N_2}n_2)|n_2\rangle\langle
n_2|,\nonumber
\end{eqnarray}
\begin{eqnarray}
V=\lambda\sum_{n_1n_2}c(n_1,n_2)\sigma^+|n_1\rangle\langle
n_2|+\texttt{H.c.},\nonumber
\end{eqnarray}
where the index $n_1$ denotes the levels of lower energy band and
$n_2$ denotes the levels of upper band, $\sigma_z$ and
$\sigma^{\pm}$ are Pauli operators. $\lambda$ is the overall
strength of the interaction, $c(n_1,n_2)$ are coupling constants,
they are independent of  each other and  are identically
distributed, satisfying
\begin{eqnarray}
\langle c(n_1,n_2)\rangle&=&0,\nonumber\\
\langle c(n_1,n_2)c(n_1',n_2')\rangle&=&0,\nonumber\\
\langle
c(n_1,n_2)c^*(n_1',n_2')\rangle&=&\delta_{n_1,n_1'}\delta_{n_2,n_2'}.\nonumber
\end{eqnarray}
According to $H_0$, one can transform the problem into the
interaction picture and, with the help of projection superoperator
technique, obtain the non-Markovian evolution equation as
\begin{eqnarray}
\frac d{dt}\rho_S^{(1)}(t)=
\gamma_1\sigma^+\rho_S^{(2)}(t)\sigma^--\frac{\gamma_2}2\{\sigma^+\sigma^-,\rho_S^{(1)}(t)\},\nonumber \\
\frac d{dt}\rho_S^{(2)}(t)=
\gamma_2\sigma^-\rho_S^{(1)}(t)\sigma^+-\frac{\gamma_1}2\{\sigma^-\sigma^+,\rho_S^{(2)}(t)\},\label{exaplemaster}
\end{eqnarray}
where
\begin{eqnarray}
\gamma_i=\frac{2\pi\lambda^2N_i}{\delta\epsilon}\quad(i=1,2).\nonumber
\end{eqnarray}
With  definitions of $\Pi_1=\sum_{n_1}|n_1\rangle\langle n_1|$ and
$\Pi_2=\sum_{n_2}|n_2\rangle\langle n_2|$, $\Pi_1+\Pi_2=I_E$, the
two non-normalized density matrixes can be obtained by
$\rho^{(i)}_S=\Tr_E(\Pi_i\rho_T),i=1,2$, where $\rho_T$ is the total
density matrix for the system and environment. The reduced density
matrix for the system is then given by
$\rho=\rho^{(1)}_S+\rho^{(2)}_S$. We note that in
Eq.(\ref{exaplemaster}), there are no environment operators other
than the two (c-number) parameters $\gamma_1,\gamma_2$. The initial
state of the environment is taken into account by means of the
distribution of initial $\rho_S^{(1)},\rho_S^{(2)}$, its effect on
the system dynamics was plotted in Figs.\ref{c1} and \ref{c2}. This
equation can be written in the form of Eq.(\ref{Gmasterequation}) by
setting  $H_i=0$, $R_{11}=R_{22}=0$,
$R_{12}=\sqrt{\gamma_1}\sigmap$, and
$R_{21}=\sqrt{\gamma_2}\sigmam$. In this model, there is only one
jump operator for each component, i.e.
$W_{12}^1=\sqrt{\gamma_1}\sigmap$ and
$W_{21}^1=\sqrt{\gamma_2}\sigmam$, and non-jump operators
$W_{mm}^0=I-i\mathcal{H}_mdt$ with
$\mathcal{H}_1=-\frac12\gamma_2\sigmap\sigmam$ and
$\mathcal{H}_2=-\frac12\gamma_1\sigmam\sigmap$.

We consider two types  of initial condition in the following
simulation. First, only the lower band of the environment is
populated, i.e. $\rho_S^{(2)}=0$. Under this condition, the reduced
system can be solved analytically. Another case is, the two bands of
the environment are all populated. With this initial condition, we
solve the master equation  numerically. In both cases, we choose an
 initial states $\ket{\phi(0)}=\ket{e}$ and
$\ket{\phi(0)}=\frac12(\ket{e}+\ket{g})$ for the system, where
$\ket{e}$ and $\ket{g}$ denote the excited state and ground state,
respectively. We compare the analytic solution and the numerical
simulation(solve the equation by Runge-Kutta method) to the results
obtained from the  quantum jump/trajectory approach in Figs.\ref{c1}
and \ref{c2}. The trajectory number in this quantum jump approach is
$\mathcal {N}=400$. We can see from the figures that the quantum
trajectory approach correctly reproduces the system evolution. The
errors are sufficiently small, although we choose a small number of
trajectories, showing that this method is efficient.

\begin{figure}
\includegraphics*[width=0.8\columnwidth,
height=0.5\columnwidth]{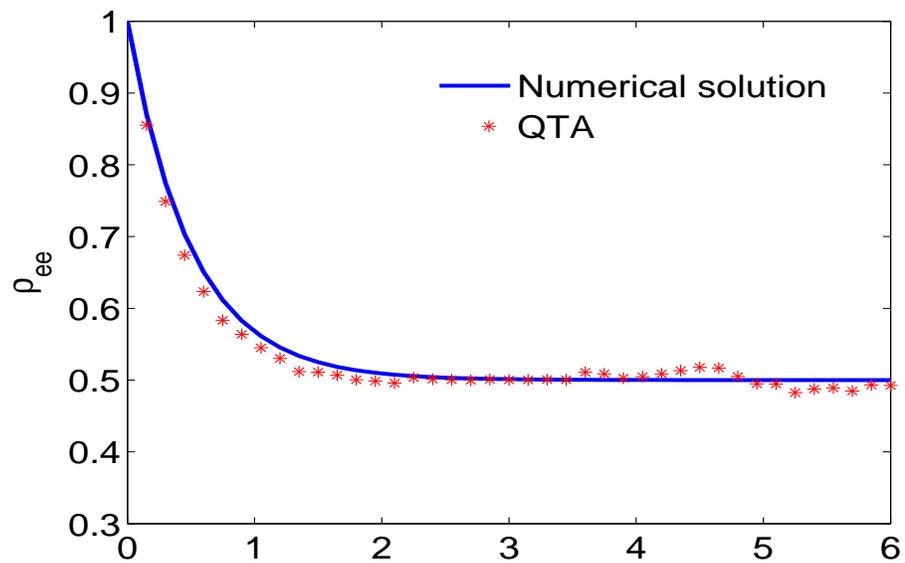}
 \caption{(Color online)
Comparisons of the numerical solution for Eq.(\ref{spinbathME}) to
the results given by quantum trajectory approach. The initial state
of the system is chosen $\rho_1=\rho_0=\rho_{-1}=\frac13\ext{e}{e}$.
All parameters in the equation are set to be equal. The time $t$ is
plotted in units of $1/\hbar$.\label{spinbathc2}}
\end{figure}

Another example is a qubit coupled to  a spin bath\cite{spinbath2}.
The full system consists of a central spin interacting with a bath
of $N$ spins. Such a system  can be described by
\begin{eqnarray}
H=\frac{\omega}2\sz+\sum_{k=1}^N\alpha_k\vec{\sigma}\cdot\vec{\sigma}_k,
\end{eqnarray}
where $\vec{\sigma}$  denotes  the Pauli matrix  for the central
spin that is the system we are interested in, and $\vec{\sigma}_k$
stands for  the $k$-th spin in the bath. After defining an
unperturbed part $H_0=\frac{\omega}2\sz+2\sz K_z$, where
$K_z=\frac12\sum_{k=1}^N\alpha_k\sz^k$, the Hamiltonian  can be
transformed into the interaction picture. Assuming the parameters
are real and time independent, the master equation reads
\begin{eqnarray}
\dt\rho_m=g_{m+1}\sigmap\rho_{m+1}\sigmam+f_{m-1}\sigmam\rho_{m-1}\sigmap
-\frac12f_m\{\sigmap\sigmam,\rho_m\}-\frac12g_m\{\sigmam\sigmap,\rho_m\},
\label{spinbathME}
\end{eqnarray}
where $\rho_m=\Tr_B(\rho_T\Pi_m)$, $\rho_T$ is the density matrix
for the total  system(the central spin plus the bath), $\Pi_m$ is a
projection superoperator that projects the $z$-component of the bath
angular momentum into an eigenvector  with eigenvalue $m$. We take
$N=2$ as an example, then the density matrix of the central spin has
three components, denoted by $\rho_1,\rho_0,\rho_{-1}$,
respectively. Each component has two jump operators which act on the
other two components, and a non-jump operator, which acts on itself.
The comparison between directly  numerical simulations (by
Runge-Kutta) and quantum trajectory method  is shown in
Fig.\ref{spinbathc2}. Here the trajectory number is chosen to be
$\mathcal{N}=4000$. We can find that as the number of jump operators
and components increases, the number of quantum trajectory, with
that we can obtain a correct result, increases.

\section{Conclusion and Discussion\label{sectionConclusion}}
In this paper, we have developed  an efficient unraveling for the
generalized Lindblad master equation. Based on this unraveling, a
generalized  Monte Carlo wave function method  is presented.  It is
worth addressing  that in this Monte Carlo wave function method, we
need only to store $M$ non-normalized wave function, i.e. $M$
length-$N$ vectors  ($M$ denotes the number of the components for
the reduced density matrix and $N$ stands for the dimension of the
Hilbert space) instead of the density operator, which are $M$
$N\times N$ matrices, hence this method saves the computer time and
space. The difference between the ordinary quantum jump method and
the present one is that the latter describes a non-Markovian
dynamics. In addition, the point that each component $\rho_i$ of the
density matrix is non-normalized, and  jumps along the component
$\rho_i$ depend on the component other than $\rho_i$ is also
different. By successfully simulating the coupling among those
components, this method can simulate the non-Markovian dynamics
efficiently. Further examination shows that the computational
complexity increases with the number of the components. The
increased complexity  due to the increase of the components and jump
operators can by analyzed as follows. Assume  the jump operators and
the number of jump operators are restricted to be the same for each
component, the possible jump mode for
$\varrho=(\rho_1,\rho_2,\cdots)$, or the number of the generalized
jump superoperators $\mathcal{W}$ is
\begin{eqnarray}
\Delta=M(J-1)+1.\label{computationalcomplexity}
\end{eqnarray}
Here  $J$ is the number of jump operators for each
component(including the non-jump operator). The role that $\Delta$
plays  is similar to the number of the jump operators in the
ordinary Markovian master equation. It is well known that one
downside  of the quantum jump approach is the complexity growth as
the jump operators proliferate. From
Eq.(\ref{computationalcomplexity}), we can find that this downside
still exists in the presented method. Still, our method is effective
when one simulates the decoherence governed by the non-Markovian
master equation, as well as for a system with Hilbert space of high
dimension.

\ \ \\

This work is supported
by  NSF of China under grant Nos. 60578014 and 10775023.\\

\newpage

\end{document}